\documentclass[lettersize,journal]{IEEEtran}
\usepackage{amsmath,amsfonts}
\usepackage{algorithmic}
\usepackage{algorithm}
\usepackage{array}
\usepackage[caption=false,font=normalsize,labelfont=sf,textfont=sf]{subfig}
\usepackage{textcomp}
\usepackage{stfloats}
\usepackage{url}
\usepackage{verbatim}
\usepackage{graphicx}
\usepackage{cite}
\usepackage{makecell}
\begin{document}

\title{Coordinated Ramp Metering Control based on Scalable Nonlinear Traffic Dynamics Model Discovery in a Large Network}

\author{Zihang Wei, Yang Zhou, \IEEEmembership{Member, IEEE}, Yunlong Zhang, Mihir Kulkarni 
\thanks{Zihang Wei, Yang Zhou, Yunlong Zhang, and Mihir Kulkarni are with the Zachry Department of Civil \& Environmental Engineering, Texas A\&M University, College Station, Texas, USA (e-mail: wzh96@tamu.edu; yangzhou295@tamu.edu; yzhang@civil.tamu.edu; mihir\_kulkarni@tamu.edu) (Corresponding Author: Yunlong Zhang)}
} 

\markboth{IEEE Transitions on Intelligent Transportation Systems}%
{Shell \MakeLowercase{\textit{et al.}}: A Sample Article Using IEEEtran.cls for IEEE Journals}


\maketitle

\begin{abstract}
This study proposes a coordinated ramp metering control framework in large networks based on scalable nonlinear traffic dynamics model discovery. Existing coordinated ramp metering control methods often require accurate traffic dynamics models in real time, however, for large-scale highway networks, since these models are always nonlinear, they are extremely challenging to obtain. To overcome this limitation, this study utilizes the Sparse Identification of Nonlinear Dynamics with Control (SINDYc) to derive the accurate nonlinear traffic dynamics model from observed data. The discovered dynamics model is then integrated into a Model Predictive Control (MPC) coordinated ramp metering controller, enabling optimized control actions that enhance traffic flow and efficiency. The proposed framework is tested on a large-scale highway network that includes three intersecting highways and eight on-ramps, which outperforms the existing approaches, demonstrating its effectiveness and potential for real-time application. This framework can offer a scalable and robust solution for improving real-time traffic management in complex urban environments.
\end{abstract}


\section{Introduction}
\IEEEPARstart{O}{ver} the past decades, the rapidly increasing vehicle demands in traffic systems have made congestion inevitable due to the limited capacity of existing highway networks. Traffic congestion incurs substantial societal costs by significantly increasing commute times, energy consumption, and crash risk. Consequently, designing traffic control measures to mitigate the negative effects of traffic congestion has emerged as a critical area of research. Various control methods have been explored, including ramp management, mainstream control, and route guidance \cite{siri2021freeway}. This study specifically focuses on traffic management through ramp metering control. We propose a data-driven model predictive control (MPC) framework for coordinated ramp metering of large-scale highway networks.

For effective traffic control measures, it is crucial to develop robust algorithms to implement control actions. Previous research has introduced various control algorithms, including feedback control \cite{papageorgiou1991alinea, smaragdis2003series, wang2014local, frejo2018feed}, optimal control and MPC \cite{hegyi2005model, bellemans2006model, zegeye2012predictive, liu2022scenario, liu2016model}, deep reinforcement learning (DRL) \cite{hu2024guided, wang2022integrated}, and hybrid models \cite{airaldi2023reinforcement, sun2024novel}.The most well-known and widely adopted algorithm is ALINEA (Asservissement Linéaire D'entrée Autoroutière) \cite{papageorgiou1991alinea}, a local feedback control algorithm that adjusts ramp metering rates based on observed downstream traffic occupancy. Several extensions of ALINEA have been developed, utilizing alternative measurements or feedback laws, such as FL-ALINEA, which is based on traffic flow, UP-ALINEA, which relies on upstream occupancy \cite{smaragdis2003series}, PI-ALINEA, which incorporates a proportional-integral (PI) controller structure \cite{wang2014local}, and Feed-Forward (FF) ALINEA, which uses a time-varying density set point \cite{frejo2018feed}, among others. While feedback control algorithms are straightforward to implement and can deliver satisfactory performance, they are often limited to a local scope, where control actions depend solely on one or a few local measurements. This approach overlooks the coordinated dynamics of a larger network with multiple ramp meters by dividing it into isolated sub-systems.

Alternatively, optimal control algorithms, such as MPC, have been applied to ramp metering. MPC is an online control algorithm that utilizes traffic dynamics models to optimize control actions within a finite prediction horizon, aiming to achieve a specified objective function. Unlike feedback control, MPC systematically optimizes the ramp metering rate for the entire network, potentially leading to optimal or sub-optimal performance \cite{wang2022integrated}. The METANET traffic network dynamics model \cite{messmer1990metanet} has been widely used to develop MPC controllers for ramp metering \cite{hegyi2005model, bellemans2006model, zegeye2012predictive}. Additionally, other dynamics models, such as FASTLANE \cite{liu2016model} and the Cell Transmission Model (CTM) \cite{gu2021smoothing}, have also been integrated with MPC for ramp metering control. The core of MPC-based ramp metering control algorithms is the traffic network dynamics model. The performance of the MPC controller heavily depends on the quality of dynamics models. These models are developed based on first principles and physical laws, which can limit the scalability of MPC-based algorithms for large highway networks, as the dynamics models become increasingly complex and almost impossible to formulate for larger networks. Moreover, real-world traffic contains uncertainties and can be affected by disturbances, making it difficult for dynamics models to accurately reflect actual traffic dynamics \cite{wang2022integrated, sun2024novel}. 

The limitations of feedback and traditional MPC algorithms are being addressed by the development of data-driven algorithms, particularly those utilizing DRL. DRL leverages the power of data to effectively address uncertainties and disturbances of highway networks. For instance, Wang et al. \cite{wang2022integrated} introduced a centralized traffic control system using DRL to coordinate ramp metering and variable speed limits on freeways, ultimately minimizing total travel time through traffic simulation. Similarly, Belletti et al. \cite{belletti2017expert} developed a DRL-based ramp metering strategy trained using the discrete LWR model. While DRL algorithms exhibit superior performance compared to feedback and Model Predictive Control (MPC) algorithms, they also have notable limitations. For example, training an effective DRL model offline requires substantial computational power, especially for large-scale networks. Additionally, DRL models are often regarded as "black boxes" due to their lack of interpretability. Moreover, unlike feedback and MPC algorithms, it is challenging to guarantee that DRL can consistently satisfy constraints on system states and control actions \cite{sun2024novel}. There are other studies that propose hybrid models that combine traditional MPC and DRL to leverage their respective advantages. Sun et al. \cite{sun2024novel} proposed a hierarchical MPC-DRL framework that used MPC for initial optimal control and DRL for fine-tuning, efficiently handling uncertainties, and improving overall control performance. Airaldi et al. \cite{airaldi2023reinforcement} developed a ramp metering strategy formulated as a reinforcement learning (RL) task, leveraging MPC as a function approximation of the optimal policy in RL. Although these works have successfully addressed the limitations of MPC (e.g., inaccurate dynamics models) and DRL (e.g., large computation time, inability to satisfy physical constraints), their performance has only been tested on a small scale, typically involving one highway and one on-ramp. 

MPC algorithms have demonstrated advantages in defining customized control objectives and constraints, optimizing systematically by considering coordinated dynamics, yielding optimal or sub-optimal control actions, and exhibiting excellent computational efficiency. However, the challenge in finding accurate dynamics models limits their application in large-scale highway networks. Although obtaining traffic dynamics models for large-scale networks based on first principles and physical laws is difficult, recent studies have shown that complex dynamic models can be accurately identified using data-driven approaches. These include dynamic mode decomposition with control (DMDc) \cite{proctor2016dynamic} and sparse identification of nonlinear dynamics with control (SINDYc) \cite{brunton2016sparse}. Moreover, the discovered dynamics models can be integrated into MPC to perform control tasks \cite{kaiser2018sparse}. DMDc is used for discovering linear dynamics models, while SINDYc is suited for nonlinear dynamics models. Studies have applied these methods to discover traffic network dynamics models \cite{wei2024discover, avila2020data}. With the development of data-driven discovery techniques of physical dynamics models, it has created opportunities for us to obtain accurate traffic dynamics models for complex traffic networks, which are challenging to obtain using traditional methods. Subsequently, the discovered dynamics models can be integrated with control algorithms such as MPC to enable effective control measures in large networks.

Since traffic network dynamics are inherently complex and nonlinear \cite{wei2024discover}, this study proposes a SINDYc-MPC framework that leverages an MPC controller, empowered by a data-driven discovered nonlinear traffic dynamics ODE model, to control ramp metering for large-scale highway networks, thereby improving overall traffic efficiency. Unlike most existing studies, which typically focus on single highways with one or several on-ramps, the proposed framework is tested on a large-scale highway network comprising three intersecting highways and eight on-ramps. The framework is designed for easy real-time implementation. Specifically, the dynamics model can be derived from data observed from the highway network where ramp metering is controlled by an existing algorithm (e.g., ALINEA). Once the dynamics model is obtained, an MPC controller can be developed with a customized objective function and constraints. Finally, the proposed framework is validated on a real traffic network simulated in SUMO (Simulation of Urban Mobility).

\section{Problem Definition}
Consider a highway network, we have a state vector $x\in \mathbb{R}^n$ represents the traffic measurements (e.g. occupancy) observed at $n$ locations within the network and a control input vector $u \in \mathbb{R}^m$ represents the metering rate at $m$ metered on-ramps to regulate traffic inflow into the highway network. There exists an ordinary differential equations (ODE) dynamics model that describes the temporal evolution of system state, $\frac{dx(t)}{dt}$, as a function of $x$ and $u$:

\begin{equation}
    \label{dynamics_ode}
    \dot{x}(t) = f(x(t),u(t)), 
\end{equation}

\noindent where $f$ is a nonlinear governing function of the dynamics, $x(t) = [x_1(t), x_2(t), ..., x_n(t)]^T$ is a state vector representing the traffic measurements observed at the $n$ traffic sensors, $\dot{x}(t)=\frac{dx}{dt}$ represents the time derivative of state vector $x$, and $u(t) = [u_1(t), u_2(t),..., u_m(t)]^T$ is a control input vector representing the metering rate at the $m$ metered on-ramps.

Based on the dynamics model in Eq.~\ref{dynamics_ode}, an MPC controller can be designed to optimize the control actions $u$ that can drive the system states $x$ to achieve a specific control objective. In this study, we consider the control objective to push traffic occupancy observed at all locations toward a desired level (i.e., optimal or capacity occupancy). According to the fundamental diagram (FD), traffic flow operating under the capacity occupancy achieves best efficiency with maximum throughput \cite{papageorgiou1991alinea, smaragdis2004flow, papageorgiou2008misapplication}. The performance of the MPC controller is highly dependent on the accuracy of the dynamics model in Eq.~\ref{dynamics_ode}. For smaller-scale networks, dynamics models are relatively easy to obtain through traffic dynamics models built according to first-principles and physical laws, such as the METANET \cite{messmer1990metanet} and the cell transmission model (CTM) \cite{daganzo1994cell}. However, for large-scale highway networks (i.e., with higher-dimensional $x$ and $u$), first-principle-based traffic network dynamics models become less accurate due to model mismatch caused by uncertainties and disturbances of traffic dynamics  \cite{siri2021freeway}, making it extremely challenging to obtain  accurate dynamics models.

\begin{figure}[!h]
\centering
\includegraphics[width=\columnwidth]{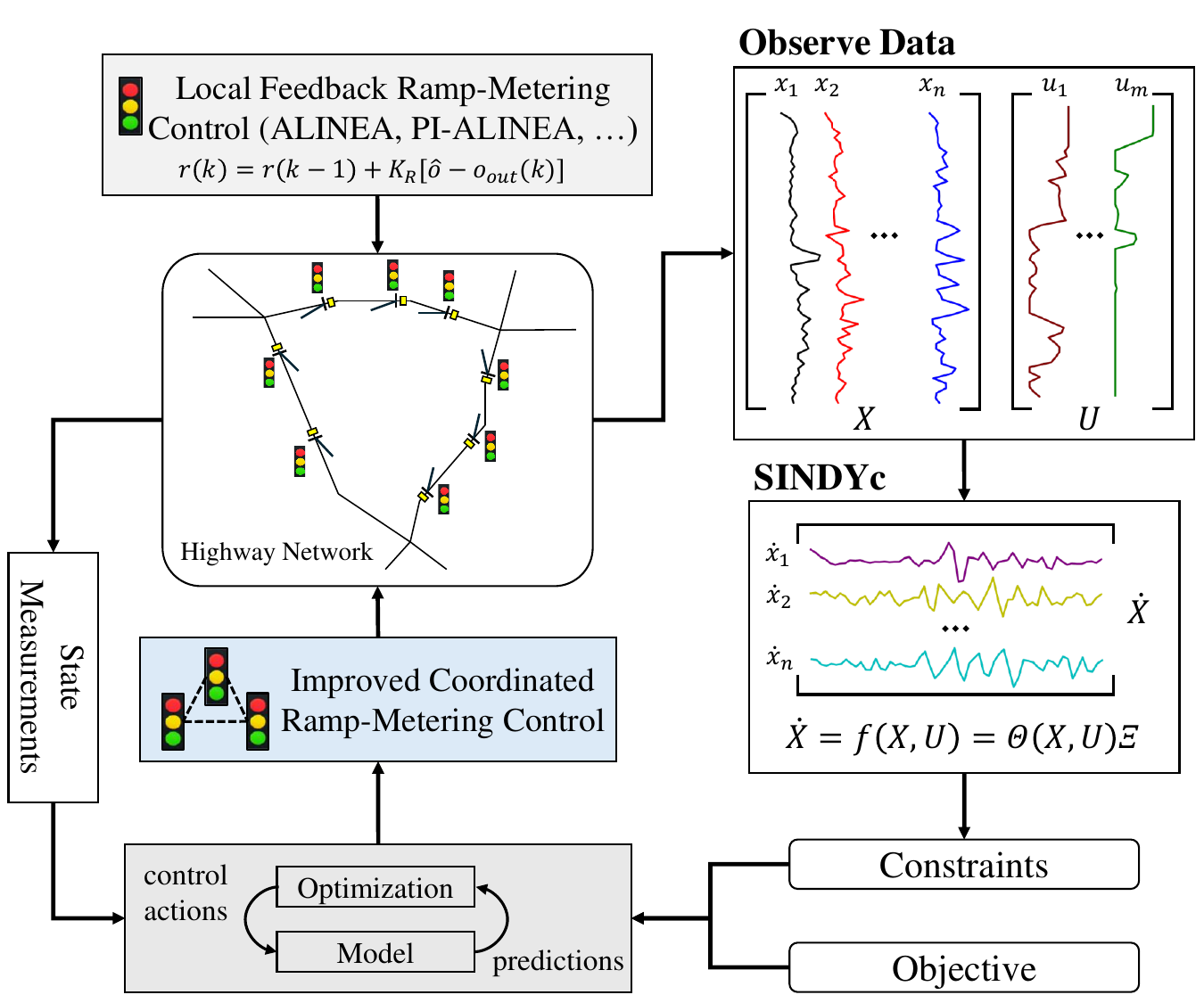}
\caption{Framework of the proposed data-driven ramp metering model predictive control.}
\label{framework}
\end{figure}

This study proposes a data-driven SINDYc-MPC framework that (1) adopts SINDYc to discover the nonlinear traffic dynamics model based on data observed in real time, and (2) designs an MPC controller based on the discovered traffic dynamics model to perform ramp metering control so that the highway network can operate in optimal efficiency. Figure \ref{framework} illustrates the high-level design of the framework.

To begin with, a local feedback control algorithm such as ALINEA and PI-ALINEA is implemented to regulate the ramp-metering rate at multiple on-ramps within the highway network. These local feedback control algorithms assume downstream traffic measurements including flow and occupancy are only influenced by the immediate upstream ramp-metering rate. However, for large-scale networks, local feedback control algorithms may not yield optimal control solutions, as they overlook the coordinate dynamics within the highway network. In order to derive the nonlinear ODE dynamics model in Eq.~\ref{dynamics_ode}, traffic occupancy data $x$ and control input data $u$ (i.e., ramp metering rate) are collected while the local feedback control algorithm is being applied to the highway network. Subsequently, the SINDYc method is used to discover the ODE dynamics model in Eq.~\ref{dynamics_ode} through a data-driven approach. Finally, a Model Predictive Control (MPC) controller with a specific objective function is designed by incorporating the discovered dynamics model into the constraints. The MPC controller can offer improved coordinated ramp metering control strategies for large-scale highway networks in a systematical perspective, thereby further enhancing operational efficiency.

\section{Methodology}
\label{sec:methodology}

\subsection{Koopman Operator Theory with Inputs and Control (KIC)}
Traffic occupancy data $x$ and control input data $u$ are collected while the local feedback control algorithm (e.g., ALINEA) is being applied to control ramp metering rate in a highway. For the state vector $x$ and the control input vector $u$, there exists an unknown nonlinear ODE dynamics model as defined in Eq.~\ref{dynamics_ode}. In this study, it is demonstrated that this unknown ODE dynamics model can be discovered from data observed in real time. First, let's consider a discrete nonlinear dynamical system of the following form:

\begin{equation}
    \label{Eq.ode_discrete}
    x(t+1) = \Tilde{f}\left(x(t)\right),
\end{equation}

\noindent where $\Tilde{f}:\mathbb{R}^{n} \mapsto \mathbb{R}^{n}$ is a function map that discretely advance the system state $x$ from step $t$ to $t+1$. Under the Koopman operator theory \cite{koopman1931hamiltonian}, there exists a observable function $g: \mathbb{R}^n \mapsto \mathcal{H}$ that maps the system state to an infinite dimensional observable Hilbert space $\mathcal{H}$ as $g(x(t))$. Then, a Koopman operator $\mathcal{K}$ directly acts on the observable Hilbert space and linearly advance $g(x(t))$ in time as:

\begin{equation}
\label{Eq.Koopman_operator}
    \mathcal{K}g(x(t)) = g(x(t+1)) = g(\Tilde{f}\left(x(t)\right)) = g \circ \Tilde{f}(x(t)),
\end{equation}

\noindent where $g \circ \Tilde{f}$ denotes the function decomposition of $g$ with $\Tilde{f}$. The Koopman operator theory is further generalized to include inputs and control\cite{proctor2018generalizing}. Consider the continuous dynamical system with control input in Eq.~\ref{dynamics_ode} as the discrete form:

\begin{equation}
    \label{Eq.ode_discrete_control}
    x(t+1) = \Tilde{f}(x(t), u(t)),
\end{equation}

\noindent where $\Tilde{f}:\mathbb{R}^{n+m} \mapsto \mathbb{R}^{n+m}$ is a function map that discretely advance the system states from step $t$ to step $t+1$. According to KIC, there also exists a observable function $g: \mathbb{R}^{n+m} \mapsto \mathcal{H}$ that acts on both the states $x$ and the control inputs $u$ which map them into an infinite dimensional observable Hilbert space $H$. The Hilbert space $\mathcal{H}$ should contains scalar value observables $g_i$ that depend on the state (i.e., $g_i(x,u)=x_1$), control inputs (i.e., $g_i(x,u)=u_1$), and mixed terms (i.e., $g_i(x,u)=x_1^2u_1^2$) \cite{proctor2018generalizing}. Then there exists a Koopman operator $\mathcal{K}:\mathcal{H} \mapsto \mathcal{H}$ acts on the observable Hilbert space as:

\begin{align}
    \mathcal{K}g(x(t), u(t)) &= g(x(t+1), u(t+1)) \nonumber\\ 
    &= g(\Tilde{f}(x(t), u(t)), u(t+1)). \label{Eq.Koopman_operator_control}
\end{align}

Essentially, the KIC defines an infinite-dimensional linear dynamical system that linearly advances the observables in the Hilbert space as $g(x(t+1), u(t+1))= \mathcal{K}g(x(t), u(t))$. Moreover, based on the type of control input, the definition of KIC in Eq.~\ref{Eq.Koopman_operator_control} can be modified. In this study, Eq.~\ref{Eq.Koopman_operator_control} is based on considering open-loop control where the control input is generated from an exogenous forcing term as a time-varying input. For other types of control, such as closed-loop control, necessary modifications of Eq.~\ref{Eq.Koopman_operator_control} can be found in section 3.1 of \cite{proctor2018generalizing}. Furthermore, if the time step $\Delta t$ defined in Eq.~\ref{Eq.Koopman_operator_control} becomes infinitely small, the discrete dynamcial system can be written as a continuous analogue:

\begin{equation}
    \label{Eq.Koopman_operator_control_continuous}
    \frac{d}{dt}g(x(t), u(t)) = \mathcal{K}'g(x(t), u(t)),
\end{equation}

\noindent where $\mathcal{K}'$ is defined as the infinitesimal generator of the one-parameter family of transformations $\mathcal{K}$ \cite{abraham2012manifolds, brunton2022data}:

\begin{equation}
    \label{Eq.Infinitesimal_generator}
    \mathcal{K}^\prime g(x(t), u(t)) = \lim_{\Delta t\rightarrow0}{\frac{\mathcal{K}g(x(t), u(t))-g(x(t), u(t))}{\Delta t}}.
\end{equation}

According to the continuous dynamical system in Eq.~\ref{Eq.Koopman_operator_control_continuous}, we can further partition the observables $g(x(t), u(t))$ in Hilbert space $\mathcal{H}$ as the following:

\begin{equation}
    \label{Eq.g_partition}
    g(x(t), u(t)) = [x(t), u(t), \varphi(x(t), u(t))],
\end{equation}

\noindent where $g(x(t), u(t))$ in $\mathcal{H}$ consists of three parts including system state $x$, control inputs $u$, and mixed terms $\varphi(x, u)$ that includes nonlinear combinations of $x$ and $u$ \cite{proctor2018generalizing}. If we denote $k(t)=\varphi(x(t), u(t))$, then Eq.~\ref{Eq.Koopman_operator_control_continuous} can be written into:

\begin{equation}
    \label{Eq.dynamics_matrix_format}
    \frac{d}{dt}g(x(t), u(t))=\begin{bmatrix}
        \dot{x}(t) \\
        \dot{u}(t) \\
        \dot{k}(t)
    \end{bmatrix} = \begin{bmatrix}
        K_{11}^\prime & K_{12}^\prime & K_{13}^\prime \\
        K_{21}^\prime & K_{22}^\prime & K_{23}^\prime \\
        K_{31}^\prime & K_{32}^\prime & K_{33}^\prime
    \end{bmatrix} \begin{bmatrix}
        x(t) \\
        u(t) \\
        k(t)
    \end{bmatrix},
\end{equation}

\noindent where $K_{11}^\prime \in \mathbb{R}^{n\times n}$, $K_{12}^\prime \in \mathbb{R}^{n\times m}$, $K_{13}^\prime \in \mathbb{R}^{n\times (h-n-m)}$, $K_{21}^\prime \in \mathbb{R}^{m\times n}$, $K_{22}^\prime \in \mathbb{R}^{m\times m}$, $K_{23}^\prime \in \mathbb{R}^{m\times (h-n-m)}$, $K_{31}^\prime \in \mathbb{R}^{(h-n-m)\times n}$, $K_{32}^\prime \in \mathbb{R}^{(h-n-m)\times m}$, and $K_{33}^\prime \in \mathbb{R}^{(h-n-m)\times (h-n-m)}$ are the coefficient matrices, where $h=\text{dim}(\mathcal{H})$. Furthermore, Eq.~\ref{Eq.dynamics_matrix_format} can be transformed into:

\begin{equation}
    \label{Eq.dynamics_matrix_format2}
    \begin{bmatrix}
        \dot{x}(t) \\
        \dot{u}(t) \\
        \dot{k}(t)
    \end{bmatrix} = \begin{bmatrix}
        K_{11}^\prime x(t) + K_{12}^\prime u(t) + K_{13}^\prime k(t)\\
        K_{21}^\prime x(t) + K_{22}^\prime u(t) + K_{23}^\prime k(t)\\
        K_{31}^\prime x(t) + K_{32}^\prime u(t) + K_{33}^\prime k(t)
    \end{bmatrix}.
\end{equation}

The nonlinear traffic dynamics model in Eq.~\ref{dynamics_ode} is correspond to the first row of Eq.~\ref{Eq.dynamics_matrix_format2}: $\dot{x}(t) = K_{11}^\prime x(t) + K_{12}^\prime u(t) + K_{13}^\prime k(t)$ and it can be discovered after identifying $k(t)$, $K_{11}^\prime$, $K_{12}^\prime$, and $K_{13}^\prime$. For the second and third row of Eq.~\ref{Eq.dynamics_matrix_format2} (i.e., $\dot{u}(t)$ and $\dot{k}(t)$), they are not directly related to the traffic network dynamics model and are not the focus of this study. However, finding the exact form of $k(t) = \varphi(x(t), u(t))$ is extremely challenging without prior knowledge of the system nonlinearity. In this study, by adopting the SINDYc method \cite{brunton2016sparse}, $k(t)$, $K_{11}^\prime$, $K_{12}^\prime$, and $K_{13}^\prime$ can be effectively approximated using a data-driven approach. 

\subsection{Sparse Identification of Nonlinear Dynamics with Control (SINDYc)}
Most physical dynamical systems are governed by only a few important forms of linearity and nonlinearity \cite{brunton2016sparse}. The SINDYc utilizes an abundant collection of linear and nonlinear terms that could potentially be related with the dynamics and uses their combination as an approximation. Firstly, based on Eq.~\ref{dynamics_ode} and \ref{Eq.dynamics_matrix_format2}, we format the traffic network dynamics model as:

\begin{equation}
    \label{Eq.dynamics_matrix_SINDYc}
    \dot{x}(t)=f(x(t), u(t))=\hat{K}_{11}^\prime x(t)+\hat{K}_{12}^\prime u(t) +  \hat{K}_{13}^\prime \hat{k}(t),
\end{equation}

\noindent where $\hat{k}(t)$, $\hat{K}_{11}$, $\hat{K}_{12}$, and $\hat{K}_{13}$ are the SINDYc approximated  $k(t)$, $K_{11}$, $K_{12}$, and $K_{13}$, respectively. Then a function map is defined as $\theta:\mathbb{R}^{n+m} \mapsto \mathbb{R}^{h}$ with $\theta(x(t), u(t))=[x(t), u(t), \hat{k}(t)]$. We then define a sparse matrix $\Xi=[\hat{K}_{11}^\prime,\hat{K}_{12}^\prime, \hat{K}_{13}^\prime]\in\mathbb{R}^{n\times h}$ by combing the columns of $\hat{K}_{11}^\prime$, $\hat{K}_{12}^\prime$, and $\hat{K}_{13}^\prime$. We can reformulate Eq.~\ref{Eq.dynamics_matrix_SINDYc} into:

\begin{equation}
    \label{Eq.dynamics_matrix_SINDYc2}
    \dot{x}(t)=f(x(t), u(t)) = \Xi \theta(x(t), u(t)),
\end{equation}

\noindent where $\theta(x(t), u(t))$ is a collection library that contains potential terms which make up the nonlinear function of $f$ in Eq.~\ref{dynamics_ode} and $\Xi$ is a sparse coefficient matrix that determines which terms are active in $\theta(x(t), u(t))$ and the corresponding coefficients of the active terms.

Next we demonstrate the process of implementing the SINDYc based on observed data of $x$ and $u$. Specifically, vector $x(t)$ and $u(t)$ are observed for a total of $d$ time steps, and these vectors can be arranged into two matrices $X \in \mathbb{R}^{d\times n}$ and $U \in \mathbb{R}^{d\times m}$ as:

\begin{align}
    X &= [x(1), x(2),\ldots, x(d)]^T. \label{X} \\ 
    U &= [u(1), u(2),\ldots, u(d)]^T. \label{U}
\end{align}

Furthermore, we define $\Theta(X,U) \in \mathbb{R}^{d\times h}$ by arranging $\theta(x(t), u(t))$ from all $d$ time steps into a matrix. Then, Eq.~\ref{Eq.dynamics_matrix_SINDYc2} can be reformatted into:

\begin{equation}
    \label{SINDy}
    \dot{X}= \Theta(X,U)\Xi^T, 
\end{equation}

\noindent where $\dot{X}=[\dot{x}(1), \dot{x}(2),\ldots, \dot{x}(d)]^T$, $\dot{x}(t)$ can be directly calculated from state vector $x$ through numerical differentiation. $\Xi$ is defined the same as in Eq.~\ref{Eq.dynamics_matrix_SINDYc2}. The design of $\Theta(X,U)$ is defined as the following:

\begin{equation}
    \label{theta_x_u}
    \Theta(X,U) = [1,X,U,(X\otimes X),(X\otimes U), (U\otimes U), \cdots],
\end{equation}

\noindent where $X\otimes U$ defines the vectors of all possible product combinations of the components in $X$ and $U$. Specifically, an example of $\Theta(X,U)$ with second-order polynomial is designed as: 

\begin{multline}
\label{theta_x_u_example}
    \Theta(X,U)=[1,\ X_1,X_n, U_1, U_m, X_1^2, X_1X_n, X_1U_1,\\ 
    X_1U_m,X_n^2, X_nU_1, X_nU_m, U_1^2, U_1U_m, U_m^2, \ldots],
\end{multline}

\noindent where $X_i, i\in [1,n]$ and $U_i, i\in[1,m]$ are the $i^{th}$ column of $X$ and $U$. For most of the dynamical systems, only a number of terms in $\Theta(X,U)$ are active, and thus, matrix $\Xi$ is mostly sparse. So, the sparse regression is applied to obtain $\Xi$:

\begin{equation}
    \label{sparse_regression}
    \xi_k = \underset{\hat{\xi}_k}{argmin}\frac{1}{2}\|\dot{X}_k - \Theta(X,U)\hat{\xi}_k\|_2^2 + \lambda\|\hat{\xi}_k\|_0,
\end{equation}

\noindent where $\dot{X}_k, k \in [1,m]$ is the $k^{\text{th}}$ column of $\dot{X}$, $\hat{\xi}_k$ is the $k^{\text{th}}$ column of $\Xi^T$, and $\lambda$ is the sparsity-promoting parameter. In this study, the sequential threshold least squares (STLS) method is applied to solve the sparse regression in Eq.~\ref{sparse_regression}. After obtaining the sparse matrix $\Xi$, the dynamics model in Eq.~\ref{Eq.dynamics_matrix_SINDYc2} is known and we can incorporate it into a MPC controller to optimize the coordinated ramp metering strategy for improving highway network operation efficiency.




\subsection{Model Predictive Control (MPC)}
MPC is an optimal control strategy where control actions are optimized over a finite horizon to minimize an objective function while satisfying several constraints. The horizontal axis represents the control step, and the vertical axis shows the ramp metering rate. The length of a control step is denoted by $t_c$, corresponds to the time interval for ramp meters to change their rates. This interval matches the control step length of the ALINEA algorithm. Each control step consists of $n$ simulation steps, where $t_c = nt_s$, with $t_s$ being the duration of each simulation step. In this study, we have $t_s = 1 \text{s}$. The prediction horizon includes a total of $N$ control steps. The MPC controller predicts the system's behavior and optimizes the control inputs over the entire prediction horizon to achieve the desired objective. At each control step, the optimization is performed over a moving prediction horizon—only the first control action is implemented, and the optimization process repeats at the next control step.

The MPC controller optimizes the control inputs to minimize the objective function. In this study, the objective function $J_k$ is designed as:

\begin{align}
    \label{objective}
    J_k = &\sum_{l=0}^{N-1}\left( \Delta x_{k+l}^TQ\Delta x_{k+l} + \Delta u_{k+l}^TR\Delta u_{k+l}\right) \nonumber\\
    & + \Delta x_{k+N}^TP\Delta x_{k+N},
\end{align}

\noindent where $k$ is the current control step, $\Delta x_{k+l}$ represents the deviation from desired occupancy $\hat{k}$ of each location at control step $k+l$ (Eq.~\ref{delta_x}), $\Delta u_{k+l}$ is the change in ramp metering rate of two consecutive control steps $k+l$ and $k+l-1$ (Eq.~\ref{delta_u}). $Q$ is the weight matrix of state deviation for intermediate control steps (Eq.~\ref{Q}), and $P$ is for the terminal control step (Eq.~\ref{P}). $R$ is the weight matrix of changes in control inputs of two consecutive control steps (Eq.~\ref{R}). Specifically, $Q$, $P$, and $R$ are designed as diagonal matrices where the values on the diagonal correspond to the $n$ elements in the state vector $x$ or the $m$ elements in the control vector $u$. 

\begin{align}
    \Delta x_{k+l} &= x(k+l) - \hat{o}. \label{delta_x} \\
    \Delta u_{k+l} &= u(k+l) - u(k+l-1). \label{delta_u} \\
    Q &= \text{diag}(q_1, q_2, \ldots, q_n). \label{Q} \\
    P &= \text{diag}(p_1, p_2, \ldots, p_n). \label{P}\\
    R &= \text{diag}(r_1,r_2,\ldots, r_m). \label{R}
\end{align}

The objective function in Eq.\ref{objective} represents the summation of deviations from the optimal occupancy $\hat{o}$ of all sensors within the prediction horizon, as well as the summations of changes in ramp metering rates during the same period. Minimizing Eq.\ref{objective} ensures that traffic occupancy remains as close as possible to the optimal value while simultaneously minimizing variations in ramp metering rates. Based on this, the MPC controller is formulated as an optimization problem as follows:

\begin{subequations}\label{opt_formulation}
\begin{align}
&\underset{u(k),...,u(k+N-1)}{\min} J_k\\
s.t.\quad &x(k) = \Tilde{x}(k) \label{cons_1}\\
&x(k+l+1) = F(x(k+l), u(k+l)), \quad l\in[0,N-1] \label{cons_2}\\
&x_{min} \leq x(k+l) \leq x_{max}, \quad l\in[1,N] \label{cons_3}\\
&u_{min} \leq u(k+l) \leq u_{max}, \quad l\in[0,N-1] \label{cons_4}
\end{align}
\end{subequations}

At each control step during the simulation (i.e., $k=0,1,2,...$), the optimization in Eq.~\ref{opt_formulation} is performed to minimize the objective function $J_k$. This will minimize, at all control steps within the prediction horizon, the total deviation of occupancy values observed at all sensor locations in the highway network. Furthermore, large variations of control inputs at 2 consecutive steps will also be penalized. Eq.~\ref{cons_1} set the initial observations of system state at current control step $k$. The system space at each step should satisfy the nonlinear traffic dynamics model (Eq.~\ref{dynamics_ode}) discovered by SINDYc as in Eq.~\ref{Eq.dynamics_matrix_SINDYc2}. Eq.~\ref{cons_2} is the discrete version of  Eq.~\ref{dynamics_ode}, which can be obtained as followed:

\begin{equation}
\begin{aligned}
    x(t+1)&=F(x(t),u(t))\\
    &=x(t) + hf(x(t),u(t)),
\end{aligned}
\end{equation}

\noindent where $F$ is a nonlinear function that discretely advances the system state, $h$ is the sampling time step and in this study, $h$ equals to 1. Moreover, Eq.~\ref{cons_3} defines the range of system state, where $x_{min}$ and $x_{max}$ are the minimum and maximum allowed occupancy values. Eq.~\ref{cons_4} defines the range of the control inputs, where $u_{min}$ and $u_{max}$ are the minimum and maximum ramp metering rates.

\section{Case Study and Model Setup}
The proposed framework is applied to a highway network simulated in SUMO \cite{krajzewicz2010traffic} for performance evaluation. In this section, we first introduce the highway network and the simulation configuration, and then model setups are presented subsequently.

\subsection{Traffic Simulation Configuration}
\begin{figure}[!h]
\centering
\includegraphics[width=0.8\columnwidth]{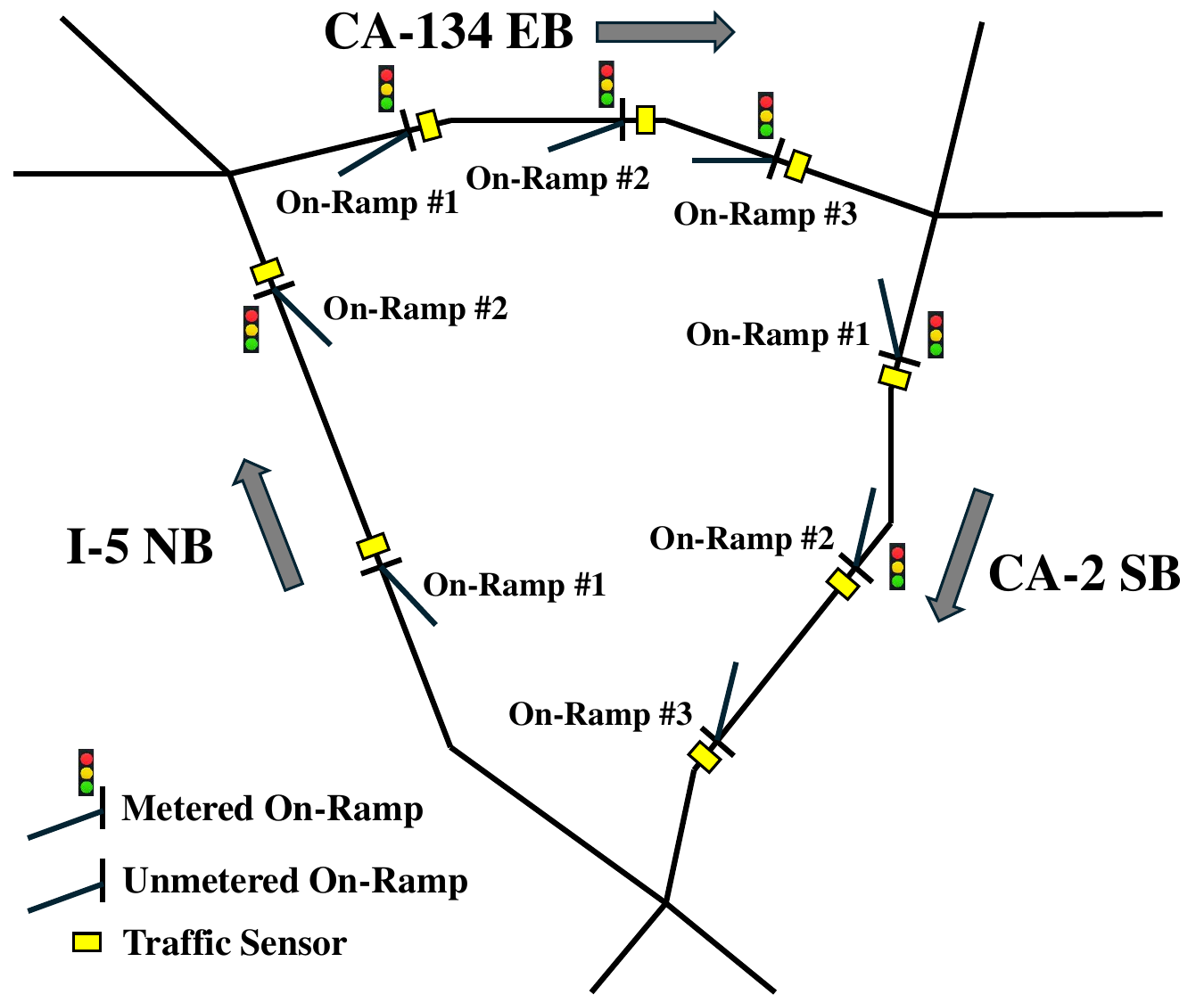}
\caption{Structure of the tested highway network including three highways: CA-134E, CA-2S, and I-5N}
\label{simulation setup}
\end{figure}

A real-world highway network surrounding the city center of Glendale, north of downtown Los Angeles,  is modeled using SUMO. This network comprises three intersecting highways: California Route 134 Eastbound (CA-134E), California Route 2 Southbound (CA-2S), and Interstate 5 Northbound (I-5N). Within the network, there are eight single-lane on-ramp meters through which traffic can enter the highway networks from local roads. Specifically, CA-134E and CA-2S each have three on-ramps, while I-5N has two. Additionally, a loop detector is installed downstream of each on-ramp to collect real-time traffic data (i.e., occupancy, flow, and speed). Fig.\ref{simulation setup} provides an illustration of the configuration of the highway network.

In SUMO, the simulation is set for a total of 1.5 hours. Since the default length of simulation step $t_s$ in SUMO is 1 second, there are 5400 simulation steps in total. During the initial simulation steps, it takes some time for the network to reach a steady state. Thus, the first 30 minutes (1800 steps) of the simulation are designed as the burn-in period to ensure the highway network reaches a steady state that can reflect real-world traffic conditions. The remaining 1 hour (3600 steps) are subjected to the control and analysis in this study. During the 1 hour simulation, the traffic demand is set to be consistently high for the necessity of ramp metering. Specifically, the demand into the network through the in-nodes of CA-134E, CA-2, and I-5N in are 3250vph, 3400vph, and 4200vph, respectively. The demand through all eight on-ramps are set to be 2000vph. Furthermore, the vehicle arrivals are set to follow Poisson distribution.

\subsection{Local Feedback Controller Setup}
This study adopts ALINEA \cite{papageorgiou1991alinea} as the local feedback controller in the first step of Fig.~\ref{framework}. ALINEA controller is formulated as: $r(k) = r(k-1) + K_R[\hat{o} - o_{out}(k)]$, where $r(k)$ represents the ramp metering rate (n vehicles per hour (vph) between control step $k$ and $k+1$, $o_{out}(k)$ is the last measured downstream occupancy (in \%, average over all lanes) between control step $k-1$ and $k$, $K_R$ = 70 veh/h/\% is the regulator parameter, and $\hat{o}$ = 15\% is the desired downstream occupancy, which is chosen by observing the shape of flow-occupancy FD \cite{papageorgiou2008misapplication}. Note that in this study, while different values of $\hat{o}$ at different locations may be more realistic, the $\hat{o}$ is set to the same values for all on-ramp locations for simplicity. Furthermore, $r(k)$ should stay in range $[r_{min}, r_{max}]$, it is truncated to $r_{min}$ = 200 veh/h or $r_{max}$ = 1800 veh/h if $r(k)$ is out of range \cite{papageorgiou2008misapplication}. The change in ramp metering rate is fulfilled by switching the ramp meter states at each on-ramp. Specifically, one metering circle has a green phase and a red phase, the green duration is fixed to 2s during which only one vehicle can pass, and the length of red phase is determined by the corresponding ramp metering rate $r$ as $M_R = (3600 - rM_G)/r$, where $r \in [200,1800]$ denotes the ramp metering rate, $M_G$ is the green duration which equals to 2s, and $M_R$ represents the red duration. The minimum value of $M_R$ is 0 when $r$ is set to 1800vph while the maximum value of $M_R$ is 16s when $r$ is set to 200vph.


\subsection{SINDYc Setup}
To set up SINDYc, an appropriate collection library, $\Theta$ is designed. In this study, we chose $\Theta$ to include polynomial terms up to the second order, as illustrated in Eq.~\ref{theta_x_u_example}. This indicates that the collection library includes a constant term, multiple linear terms, and multiple quadratic terms. In fact, more complex terms such as higher-order polynomial terms, and sine and cosine terms can be added to $\Theta$. This could potentially increase the accuracy of the discovered dynamics models; however, it would also increase the computation resources required. Furthermore, we set the sparsity-promoting parameter $\lambda$ in Eq.~\ref{sparse_regression} to 0.05. For the STLS algorithm that solves Eq.~\ref{sparse_regression}, we set the threshold to 0.0002.

\subsection{MPC Setup}
The MPC controller utilizes several parameters, as detailed in Table~\ref{MPC_Parameters}. The diagonal elements of the weight matrices $Q$ and $P$ are set to 1. For simplicity, the diagonal elements of the weight matrix $R$ are set to 0. The desired occupancy value $\hat{o}$ is set as the same with ALINEA. The minimum occupancy, $x_{min}$, is set to 0\%. The maximum occupancy, $x_{max}$, is set to 80\%, as occupancy larger than this is generally not possible. The minimum and maximum ramp metering rates ($r_{min}$ and $r_{max}$) are aligned with those of ALINEA, set at 200 vph and 1800 vph, respectively.

\begin{table}[!h]
\small
\caption{MPC Parameters}
\label{MPC_Parameters}
\centering
\begin{tabular}{ccc}
\hline
\textbf{Notation} & \textbf{Definition} & \textbf{Value} \\ \hline
$q_1,\ldots,q_n$ & Weights in matrix $Q$ & 1\\
$p_1,\ldots,p_n$ & Weights in matrix $P$ & 1\\
$r_1,\ldots,r_n$ & Weights in matrix $R$ & 0\\
$\hat{o}$ & Desired occupancy & 15\%\\
$N$ & Prediction horizon & 4\\
$x_{min}$ & Min allowed traffic occupancy & 0\%\\
$x_{max}$ & Max allowed traffic occupancy & 80\%\\ 
$u_{min}$ & Min ramp metering rate & 200vph \\ 
$u_{max}$ & Max ramp metering rate & 1800vph\\ \hline
\end{tabular}
\end{table}

The prediction horizon $N$ is closely related to the performance of the MPC controller. A shorter prediction horizon leads to efficient computation, while it may harm the accuracy if more future disturbances need to be considered. In contrast, a longer prediction horizon results in greater computational complexity, while the accuracy can be increased since more future events are considered. However, if the dynamics model is not accurate enough, longer prediction horizons could amplify the errors and result in less optimal results. We test the MPC models with $N$ ranging from 3 to 7 and based on the experiment results, $N$ = 4 yields the best performance in terms of increasing traffic throughput observed at all sensors and thus, we select the MPC model with $N$ = 4 as the proposed model. 


\section{Results and Discussions}
\subsection{Control Results of Local Feedback Controller}
\begin{figure}[!h]
\centering
\includegraphics[width=\columnwidth]{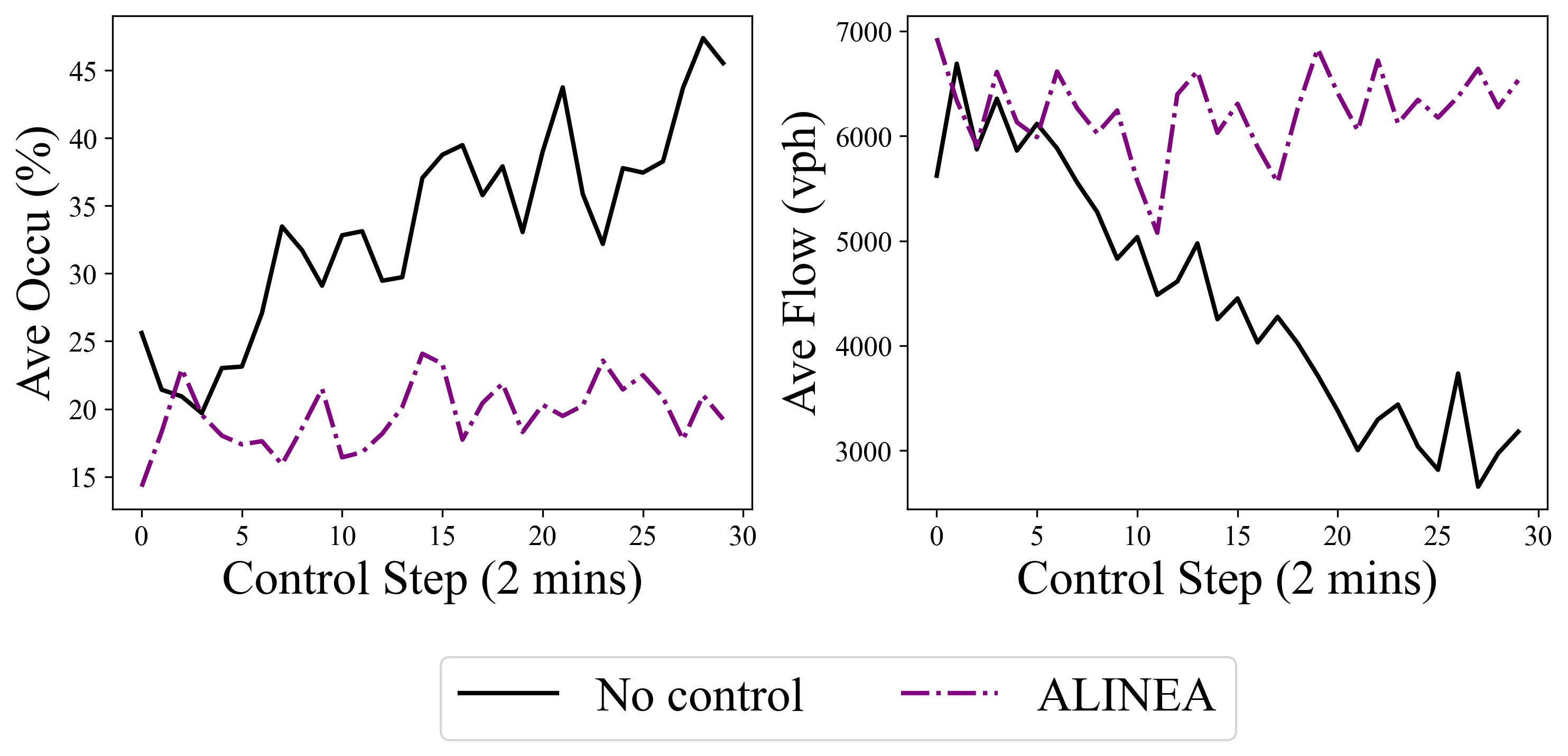}
\caption{Comparison of average traffic occupancy and flow between the no-control case and the local feedback control case.}
\label{Control_Bench}
\end{figure}

Before implementing the proposed SINDYc-MPC framework, the local feedback algorithm is first adopted to control ramp metering in the highway network for data collection. Fig.\ref{Control_Bench} presents the average traffic measurements from all eight sensors, compared with the measurements observed when ramps are not metered. Specifically, when no control measures are applied to the on-ramps, significant congestion and delays are observed in the highway network, characterized by increased occupancy as well as decreased flow over time. In contrast, when ALINEA is applied to control the ramp metering rate, the average occupancy is maintained at a lower level, and the decrease in average flow is not observed. This indicates that ramp metering control is necessary for the simulated highway network under the given traffic demand conditions to reduce congestion and improve efficiency.

\subsection{Traffic Dynamics Model Discovered by SINDYc}
\begin{figure}[!h]
\centering
\includegraphics[width=\columnwidth]{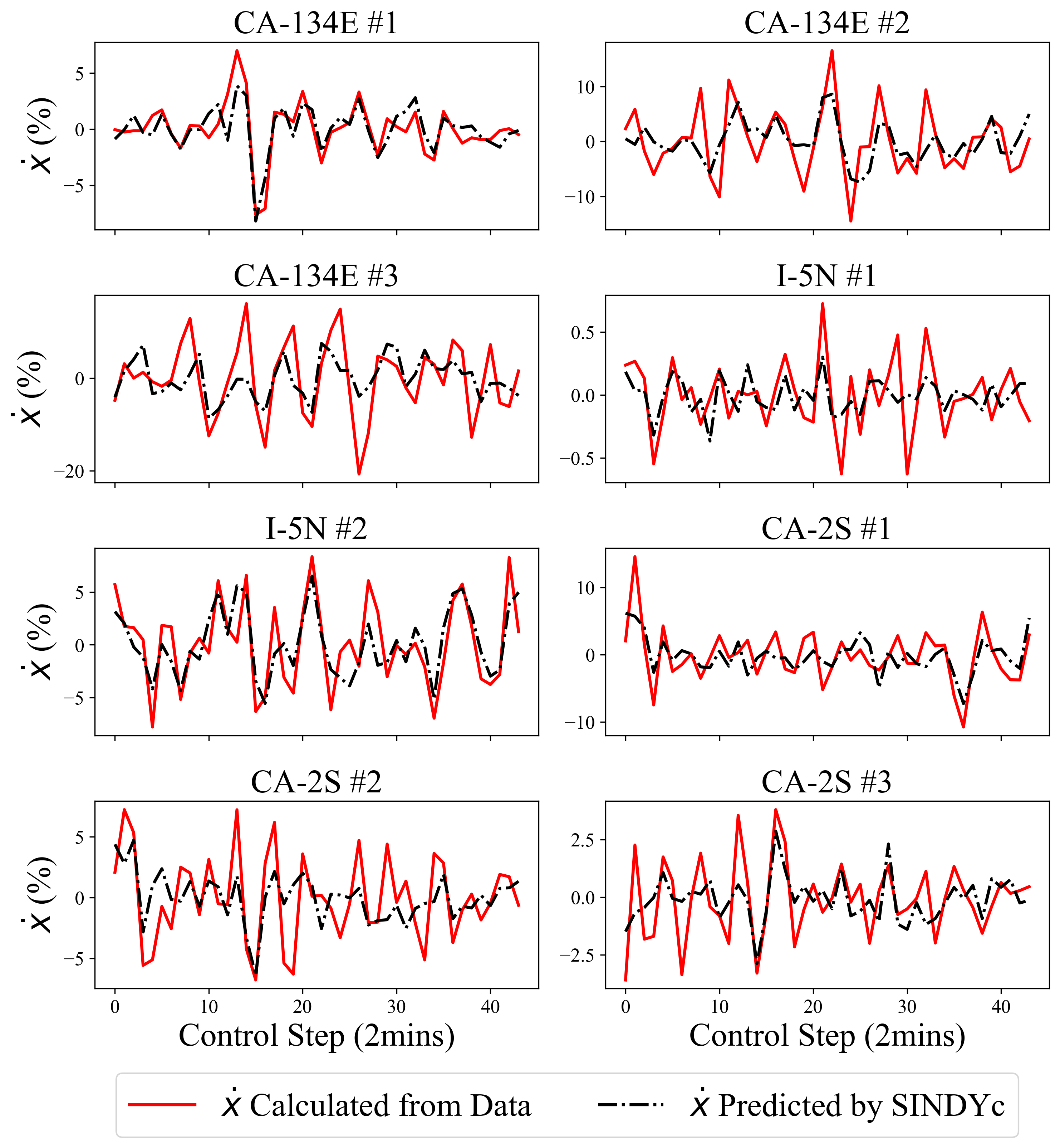}
\caption{Comparison between $\dot{x}$ predicted by SINDYc and $\dot{x}$ through numerical differentiation}
\label{SINDYc_Model_Result}
\end{figure}

While applying local feedback ramp metering control, at each control step $k$, the current traffic occupancy at all sensors is observed and stored in the state vector $x(k)$, where the current occupancy at sensor $i$, $x_i(k)$, is defined as the average occupancy over all simulation steps within the control steps $k-1$ and $k$. Furthermore, the ramp metering rates applied at step $k$ are stored in the control input vector $u(k)$. For ramp meter $i$, the metering rate $u_i(k)$ is a constant during control step $k$ and $k+1$. The vectors $x$ and $u$ at all control steps are arranged into matrices $X$ and $U$, as shown in Eq.~\ref{X} and \ref{U}, and SINDYc is applied to discover the traffic dynamics ODE model. Fig.\ref{SINDYc_Model_Result} presents the $\dot{x}(k)$ directly calculated from data $x(k)$ through numerical differentiation, alongside the $\dot{x}(k)$ calculated by the traffic dynamics model $\dot{x}(k)=\hat{f}(x(k),u(k))$ discovered by SINDYc. As presented in Fig.\ref{SINDYc_Model_Result}, the discovered traffic dynamics model can accurately compute the time derivative of traffic occupancy $\dot{x}$. This model is then incorporate into the constraint of the MPC controller to control ramp metering coordinately (i.e., Eq~\ref{cons_2}).

\subsection{Control Results of the Proposed SINDYc-MPC Framework}
The discovered traffic dynamics model is incorporated into an MPC controller and the SINDYc-MPC model (Eq.~\ref{opt_formulation}) is applied to coordinately control the ramp metering rates in the highway network. To compare with the performance of the proposed model, this study selects ALINEA, its proportional-integral extension (PI-ALINEA) \cite{wang2014local}, and a linear MPC controller based on a linear dynamics model discovered by dynamics mode decomposition (DMD-MPC) \cite{otto2024dynamic} as benchmark models. 

Table~\ref{Occu_table} presents the average occupancy deviations from the desired occupancy $\hat{o}$, measured as $\left| x - \hat{o} \right|$, for each traffic sensor under the no-control scenario, ALINEA, PI-ALINEA, DMD-MPC, and the proposed SINDYc-MPC model. The minimum deviation for each sensor is highlighted in bold. Compared to the no-control scenario, all ramp metering strategies effectively reduce deviations from the target occupancy $\hat{o}$. Notably, the SINDYc-MPC model outperforms the benchmark models in driving average occupancy towards $\hat{o}$. Across the eight traffic sensors, the SINDYc-MPC achieves the lowest occupancy deviation at four sensors and the lowest average deviation overall, with a reduction to 7.51\%.

\begin{table}[!h]
\caption{Average Deviation from Desired Occupancy $\hat{o}$ (\%) Observed at each Sensor}
\label{Occu_table}
\centering
\begin{tabular}{lccccc}
\hline
            & No Control & ALINEA & PI-ALINEA & DMD-MPC & Proposed \\ \hline
134E \#1 & 18.37       & \textbf{5.1}    & 6.38      & 7.05    & 5.73     \\
134E \#2 & 54.25       & 10.69  & 14.11     & 16.95   & \textbf{8.75}     \\
134E \#3 & 20.85       & 8.28   & 7.23      & \textbf{4.86}    & 6.03     \\
5N \#1    & 10.47       & \textbf{6.66}   & 6.69      & 6.72    & 6.67     \\
5N \#2    & 11.81       & \textbf{5.16}   & 5.56      & 5.49    & 5.88     \\
2S \#1   & 25.54       & 11.65  & 13.63     & 11.23   & \textbf{10.21}    \\
2S \#2   & 20.08       & 16.79  & 18.83     & 16.88   & \textbf{15.26}    \\
2S \#3   & 3.71        & 2.13   & 2.85      & 2.47    & \textbf{1.57}     \\
Average     & 20.64       & 8.31   & 9.41      & 8.96    & \textbf{7.51}    \\ \hline
\end{tabular}
\end{table}

Furthermore, Table~\ref{Flow_table} presents the average traffic flow improvements observed at each sensor for the ALINEA, PI-ALINEA, DMD-MPC, and the proposed SINDYc-MPC models, relative to the no-control scenario. For each sensor, the highest traffic flow improvement is highlighted in bold. The proposed framework achieves the greatest traffic flow increase across all eight traffic sensors, with an overall average improvement of 1999 vph compared to the no-control case. This corresponds to a 11.43\% increase over ALINEA, a 31.69\% increase over PI-ALINEA, and a 14.89\% increase over DMD-MPC.

\begin{table}[!h]
\caption{Average Traffic Flow Improvement (vph) Compared with No Control at each Sensor}
\label{Flow_table}
\centering
\begin{tabular}{lcccccccc}
\hline
           & ALIANA & PI-ALINEA & DMD-MPC & Proposed  \\
134E \#1 & 3000   & 2638      & 2844    & \textbf{3157}     \\
134E \#2 & 3289   & 2876      & 3175    & \textbf{3526}     \\
134E \#3 & 3519   & 2959      & 3374    & \textbf{3883}     \\
5N \#1    & 798    & 791       & 779     & \textbf{821}      \\
5N \#2    & 1627   & 1665      & 1601    & \textbf{1730}     \\
2S \#1   & 1134   & 758       & 1068    & \textbf{1382}     \\
2S \#2   & 537    & 245       & 559     & \textbf{797}      \\
2S \#3   & 447    & 209       & 516     & \textbf{694}      \\
Average     & 1794   & 1518      & 1740    & \textbf{1999}     \\ \hline
\end{tabular}
\end{table}

\begin{figure}[!h]
\centering
\includegraphics[width=\columnwidth]{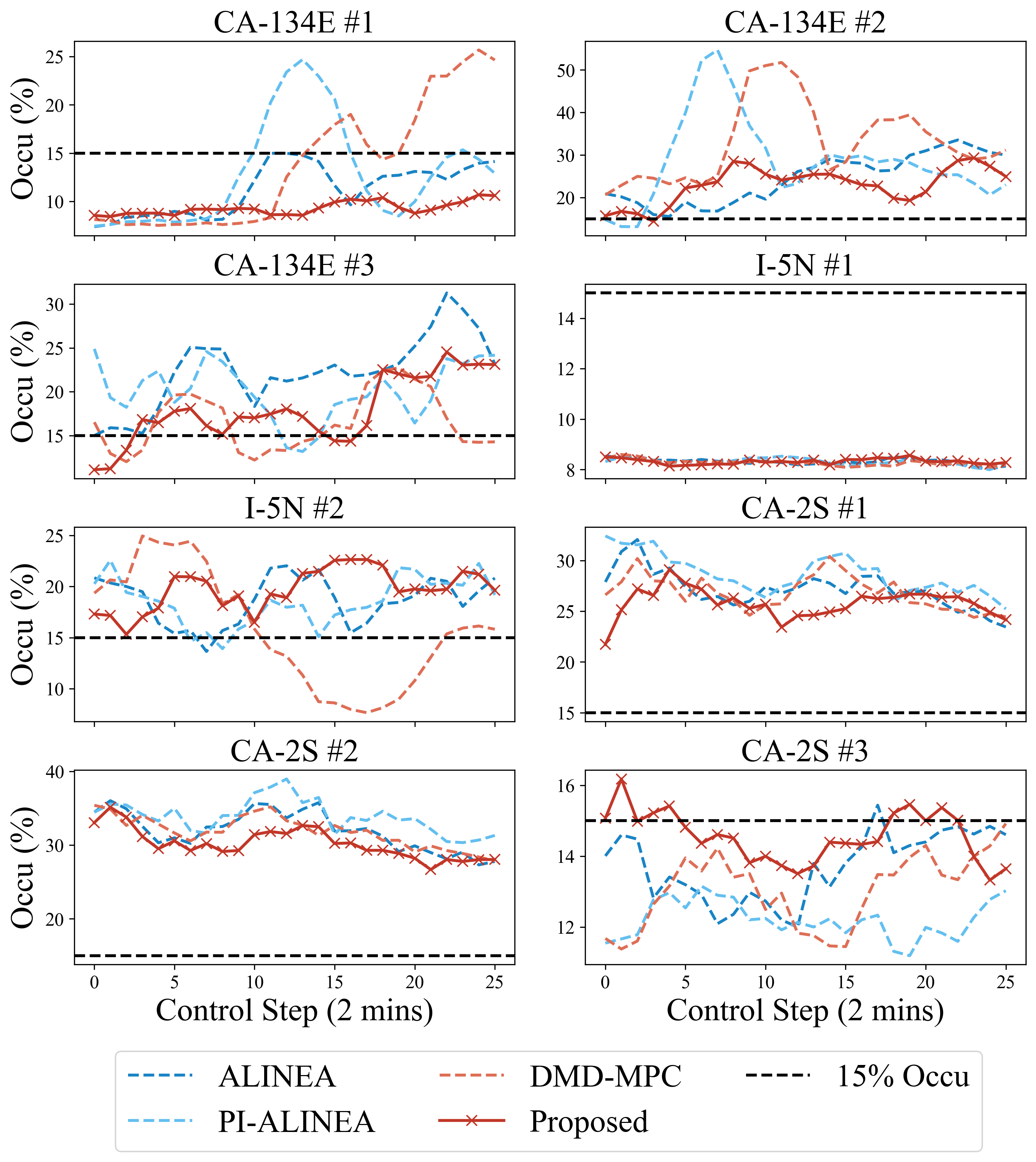}
\caption{Traffic Occupancy (\%) at each sensors of the proposed and benchmark models (moving average is applied to smooth the lines).}
\label{Occu_results}
\end{figure}

\begin{figure}[!h]
\centering
\includegraphics[width=\columnwidth]{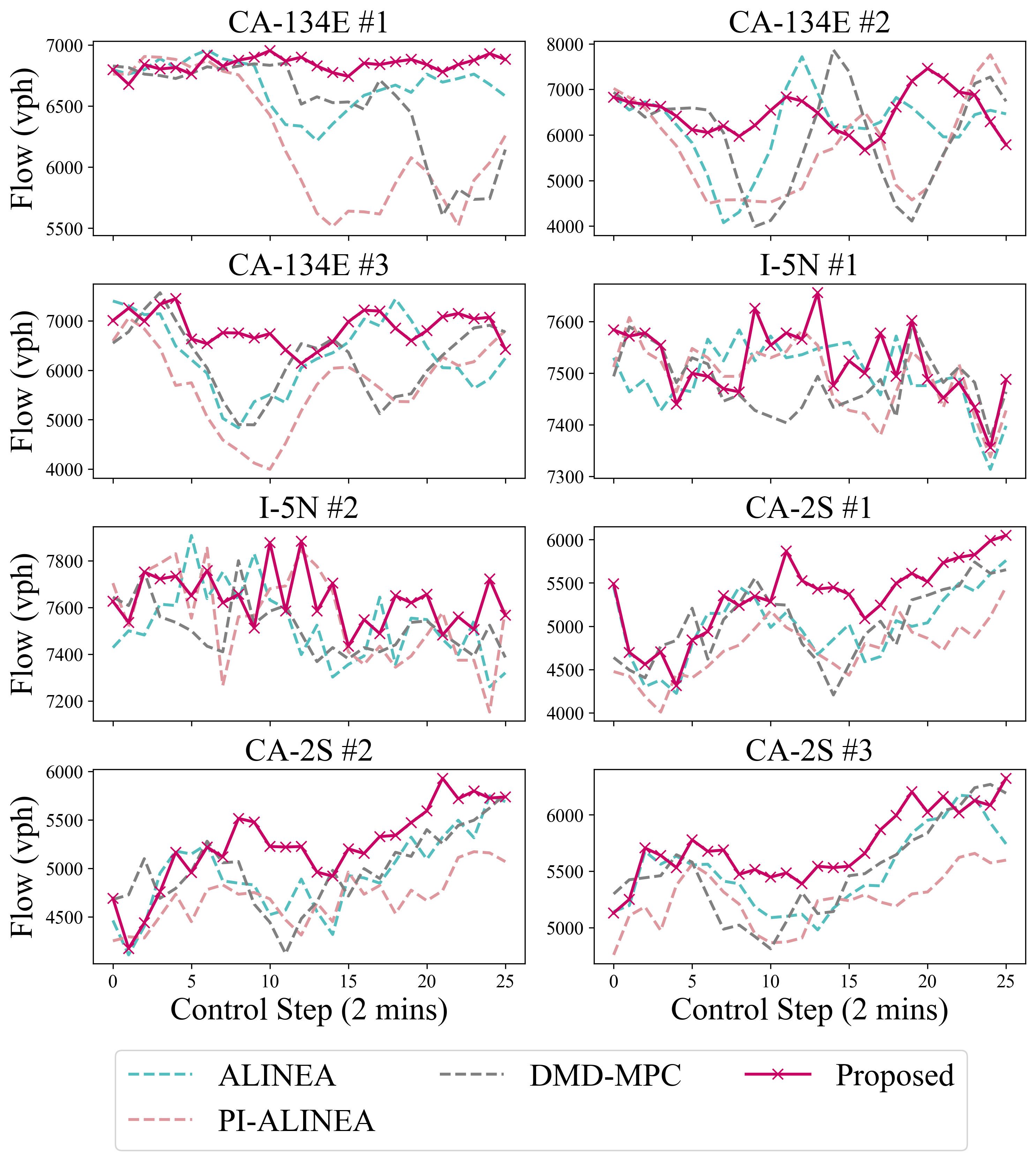}
\caption{Traffic Flow (vph) at each sensors of the proposed and benchmark models (moving average is applied to smooth the lines).}
\label{Flow_results}
\end{figure}

Moreover, the traffic occupancy observed at eight sensors resulted by ALINEA, PI-ALINEA, DMD-MPC, and the proposed model are illustrated in Fig.~\ref{Occu_results}, where the horizontal dashed line is the reference line for $\hat{o}$ (15\%). In general, comparing with the benchmark models, the proposed model can effectively keep the occupancy observed at eight sensors closer to the desired 15\%. Similarly, the traffic flow observed at eight sensors are presented in Fig.~\ref{Flow_results}. For all eight sensors, the traffic flow values observed for the proposed model are higher than those of the benchmark models.

\begin{figure*}[!h]
\centering
\includegraphics[width=\textwidth]{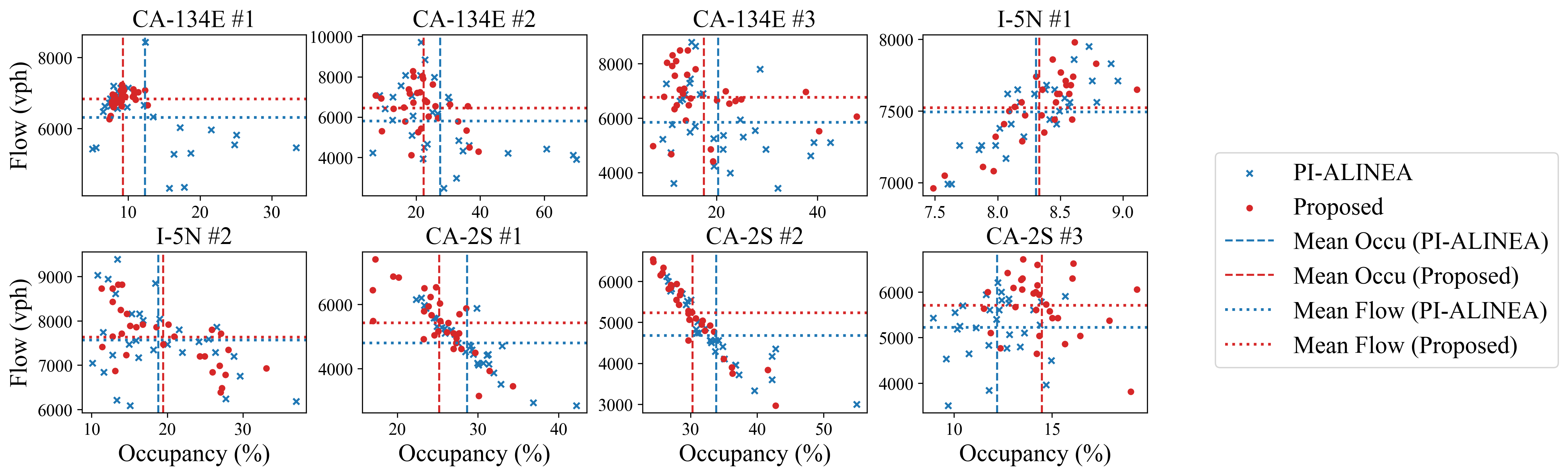}
\caption{Scatter plots of occupancy against flow for the proposed model and PI-ALINEA.}
\label{scatter_FD}
\end{figure*}

Furthermore, Fig.~\ref{scatter_FD} presents the scatter plots of traffic occupancy versus flow observed at four selected traffic sensors for the PI-ALINEA and the proposed SINDYc-MPC framework. For all sensors except ``CA-134E \#1" and ``I-5N \#2", when the average occupancy is closer to the desired occupancy $\hat{o}$ (15\%), the average flow is higher. The cases of ``CA-134E \#1" and ``I-5N \#2" suggest that the desired occupancy at these two locations are not 15\%. For instance, the shape of the fundamental diagram (FD) of sensor ``CA-134E \#1" suggests that the desired occupancy $\hat{o}$ is not 15\%, as the traffic flow decreases when occupancy increases from 10\% to 15\%. Thus, even though the proposed model drives the average occupancy away from 15\%, the average flow remains higher. The effects of the proposed model in driving the traffic occupancy closer to $\hat{o}$ (15\%) to increase traffic flow are most obvious for sensors including ``CA-134E \#2", ``CA-134E \#3", ``CA-2S \#1", ``CA-2S \#2", and ``CA-2S \#3". Furthermore, for ``I-5N \#1", the observed data points are located in the free flow region of the FD, as the proposed model increases the average occupancy, the average flow also increases. For ``CA-2S \#1" and ``CA-2S \#2", the observed data points are located in the congested region of the FD, the proposed model decreases the average occupancy and the average flow increases. As for ``CA-2S \#3", the proposed drives the average occupancy around 15\%, and the average flow is higher than that of PI-ALINEA. Based on the observations in Fig.~\ref{scatter_FD}, we can infer that the desired occupancy values may vary across different locations. While this study assumes a uniform desired occupancy $\hat{o}$ across all locations and shows that the proposed framework effectively enhances traffic efficiency, future studies could improve the framework's performance by setting different $\hat{o}$ values for different locations and times.

Finally, the average green percentage for each metered on-ramp throughout the simulation using ALINEA, PI-ALINEA, DMD-MPC, and the proposed model are summarized in Table~\ref{average_ramp_rate}. A higher green percentage signifies reduced average waiting times and delays experienced by on-ramp traffic, as vehicles are discharged into the network more efficiently. Among the models compared, the proposed model achieves the highest average green percentage. At the same time, the proposed model demonstrates the greatest improvement in average traffic flow in the main highway network. This highlights the effectiveness of the proposed framework, as it not only maximizes traffic flow on the main highway but also significantly reduces delays at the on-ramps.

\begin{table}[!h]
\caption{Average Green Percentage at each Metered On-Ramp}
\label{average_ramp_rate}
\centering
\begin{tabular}{lcccc}
\hline
                    & ALINEA & PI-ALINEA & DMD-MPC & Proposed \\
134E OR \#1 & 96.8\% & 81.0\%    & \textbf{97.6\%}  & 93.2\%         \\
134E OR \#2 & 17.7\% & 19.3\%    & 23.1\%  & \textbf{23.5\%}         \\
134E OR \#3 & 38.4\% & 23.6\%    & \textbf{58.2\%}  & 55.8\%         \\
5N OR \#2    & 19.2\% & 20.7\%    & 12.1\%  & \textbf{25.7\%}         \\
2S OR \#1   & 18.9\% & 17.9\%    & 14.1\%  & \textbf{24.9\%}         \\
2S OR \#2   & 11.1\% & 11.1\%    & 11.1\%  & \textbf{12.3\%}         \\
Average        & 33.7\% & 28.9\%    & 36.0\%  & \textbf{39.2\%}        \\ \hline
\end{tabular}
\end{table}

\section{Conclusion}
Traffic congestion can significantly impact the operational efficiency of highway networks. Traffic control measures, including ramp metering, can effectively mitigate the negative effects of increasing traffic demands. Previous studies have proposed various control algorithms for implementing ramp metering strategies in highway networks. These algorithms can be broadly categorized into three groups: local feedback control, Model Predictive Control (MPC), and Deep Reinforcement Learning (DRL). While each of these algorithms is capable of performing ramp metering effectively, they also have limitations. Local Feedback control may not account for coordinated dynamics; MPC is typically limited to small-scale networks due to the complexity of the dynamics model; and DRL, often considered a "black box," is prohibitive to interpret and requires substantial data and computational power.

MPC algorithms have demonstrated advantages in terms of computation efficiency, flexibility in designing custom control objectives and constraints, and in considering coordinated dynamics, often yielding optimal or sub-optimal control actions. However, for large-scale highway networks, the core of an MPC controller—the dynamics model—is impossible to obtain based solely on first principles and physical laws. To address this limitation, this study proposes a data-driven MPC framework, named SINDYc-MPC, to systematically control ramp metering in a large-scale highway network based on a dynamics model discovered through data-driven methods. Specifically, traffic occupancy data and ramp metering rate data are collected in real-time while the commonly used local feedback control algorithms (e.g., ALINEA, PI-ALINEA, FL-ALINEA) are being applied. Then, the SINDYc model is used to discover the nonlinear traffic dynamics model, which is later incorporated into the constraints of the MPC controller to control ramp metering in a real highway network using the traffic simulation tool SUMO. Notably, unlike previous studies validated only on relatively small networks, the proposed framework is tested on a large-scale network with three intersecting highways and eight on-ramps. Simulation results demonstrate that compared to ALINEA, PI-ALINEA, and DMD-MPC, the proposed framework can improve operational efficiency, characterized by increased traffic throughput in the main highway network, and simultaneously reduce the delay for vehicles waiting on ramps, as indicated by higher average ramp metering rates.

The proposed framework is easier to implement, as data can be collected in real-time when a common ramp metering control algorithm is applied, and it requires significantly less training time compared to other data-driven algorithms (e.g., DRL). Several further study directions can be considered. Firstly, the desired occupancy should be set to different values at different locations. Secondly, more advanced data-driven dynamics model discovery techniques can be applied. Last but not least, different MPC setups should be examined to achieve other more complicated control objectives.

\bibliographystyle{IEEEtran}
\bibliography{references}

\vspace{-1.5em}
\begin{IEEEbiography}[{\includegraphics[width=1in,height=1.25in,clip,keepaspectratio]{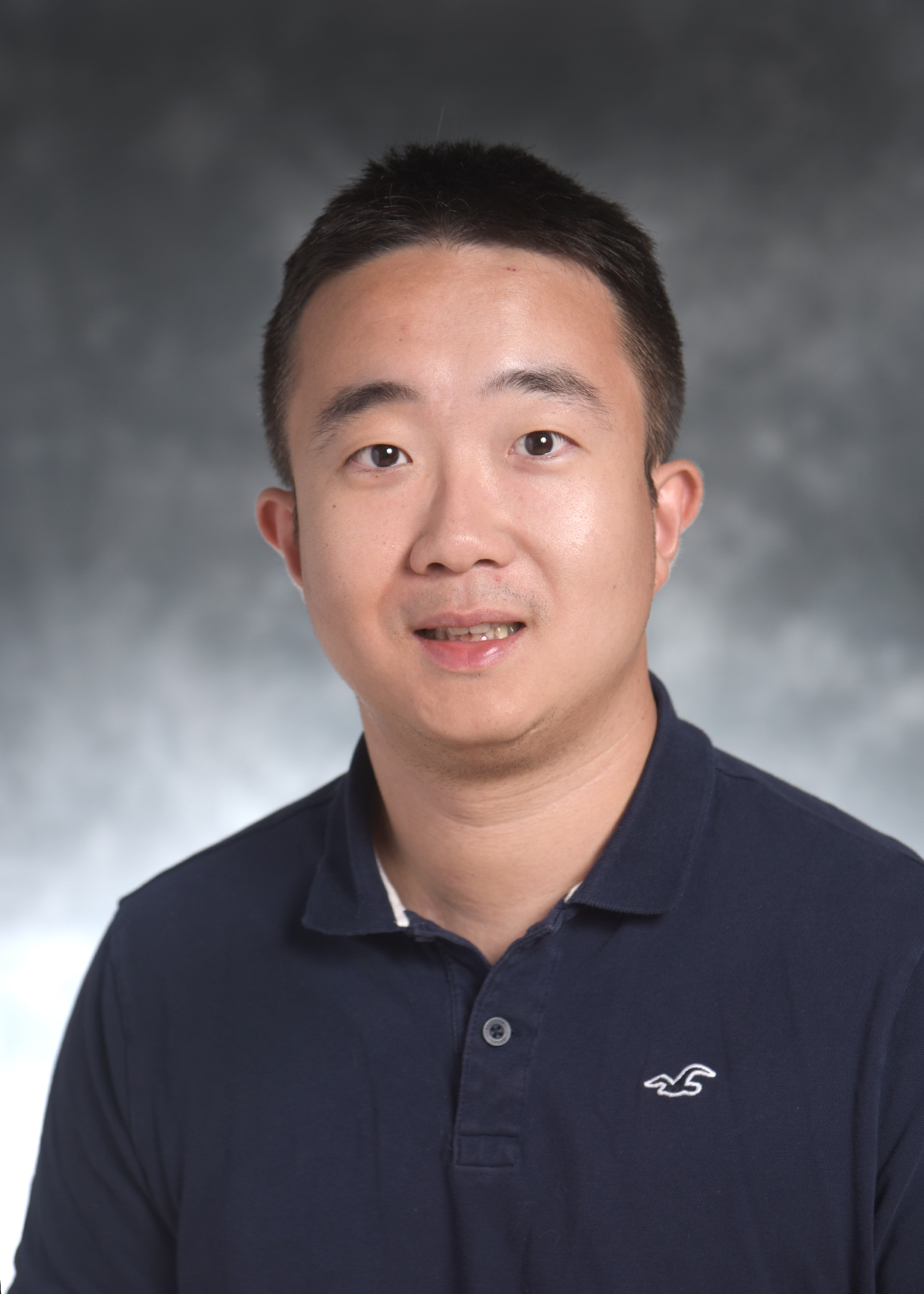}}]{Zihang Wei}received his M.Sc. degree in Civil Engineering from Texas A\&M University, College Station, TX, USA, in 2021. He is currently a Ph.D. candidate at the Zachry Department of Civil \& Environmental Engineering, Texas A\&M University. He is also working as a graduate research assistant at Texas A\&M Transportation Institute (TTI). His research interests include physics-informed modeling of traffic dynamics, traffic network control, equity in transportation systems, and traffic safety data analysis.
\end{IEEEbiography}
\vspace{-1.5em}
\begin{IEEEbiography}
[{\includegraphics[width=1in,height=1.25in,clip,keepaspectratio]{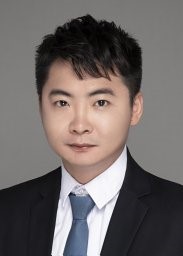}}]{Yang Zhou} received the Ph.D. degree in Civil and Environmental Engineering from University of Wisconsin Madison, WI, USA, in 2019, and the M.S. degree in Civil and Environmental Engineering from University of Illinois at Urbana-Champaign, Champaign, IL, USA, in 2015. He is currently an Assistant Professor in the Zachry Department of Civil and Environmental Engineering at Texas A\&M University. Before joining Texas A\&MUniversity, he was a postdoctoral researcher in Civil Engineering, University of Wisconsin Madison, WI, USA. He is currently a member in TRB traffic flow theory CAV subcommittee, network modeling CAV subcommittee, and American Society of Civil Engineering. His main research directions are connected automated vehicle robust control, interconnected system stability analysis, traffic big data analysis, and microscopic traffic flow modeling.
\end{IEEEbiography}
\vspace{-1.5em}
\begin{IEEEbiography}
[{\includegraphics[width=1in,height=1.25in,clip,keepaspectratio]{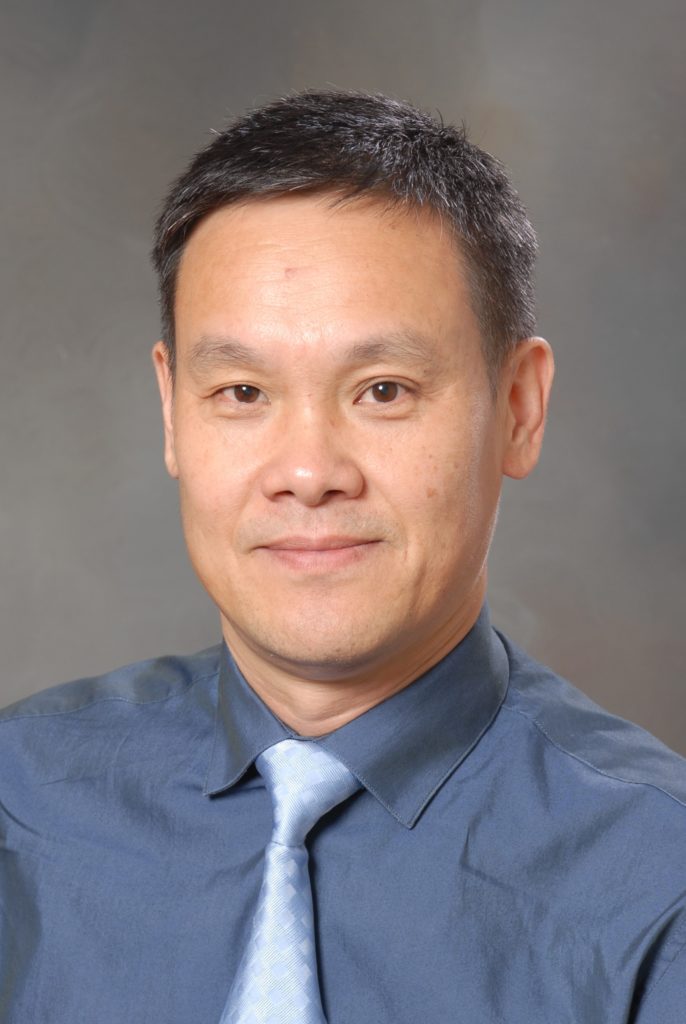}}]{Yunlong Zhang}received the B.S. degree in civil engineering and the M.Sc. degree in highway and traffic engineering from the Southeast University of China, Nanjing, China, in 1984 and 1987, respectively, and the Ph.D. degree in transportation engineering from the Virginia Polytechnic Institute, State University, Blacksburg, VA, USA, in 1996. He is currently a professor with the Zachry Department of Civil \& Engineering, Texas A\&M University, College Station, TX, USA. His research interests include transportation modeling and simulation, traffic operations, and Artificial Intelligence and data analytics applications in transportation.
\end{IEEEbiography}
\vspace{-1.5em}
\begin{IEEEbiography}
[{\includegraphics[width=1in,height=1.25in,clip,keepaspectratio]{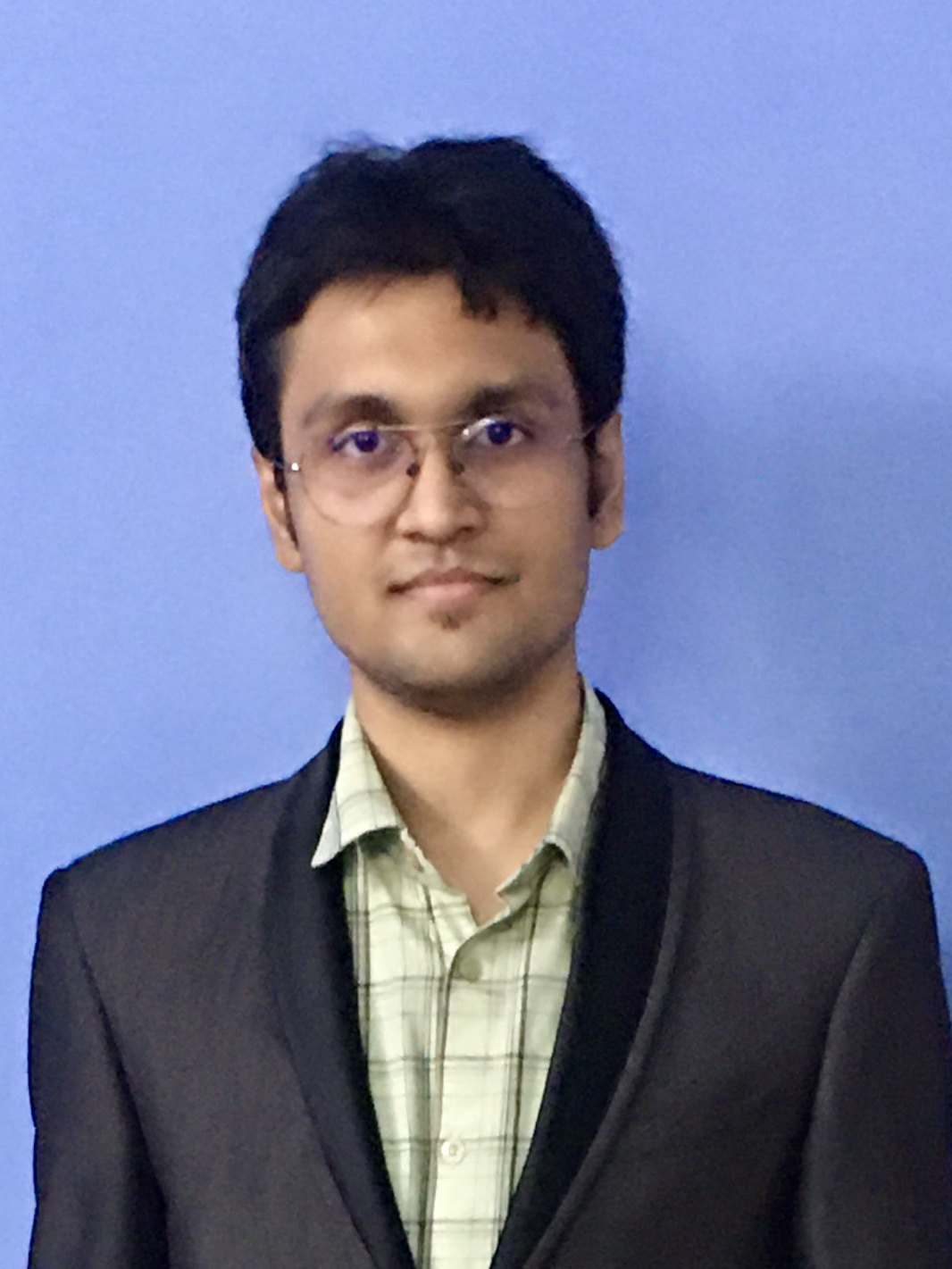}}]{Mihir Kulkarni} received B.Tech in Civil Engineering and M.Tech in Transportation Engineering from Indian Institute of Technology Madras, India in 2022. He is currently a Ph.D. student in the Zachry Department of Civil \& Environmental Engineering at Texas A\&M University. Specialized in Transportation Engineering, he is working as a research assistant at Texas A\&M Transportation Institute (TTI). His research interests are traffic modeling of autonomous vehicles, lane-changing models, and applications of connected vehicles.
\end{IEEEbiography}
\vfill
\end{document}